\documentclass[a4paper,12pt]{article}
\usepackage[reqno,fleqn]{amsmath}
\usepackage{color,graphicx}
\usepackage[T1]{fontenc}
\author{Elnaz Amirkhanlou\footnote{eliamirkhanlou@yahoo.com}, Behnam Mohammadi\footnote{be.mohammadi@urmia.ac.ir}\\
Department of Physics, Urmia University, Urmia, Iran}
\title{Contributions of $\psi_{2}(3823)$ and $\psi(4040)$ charmonium in $B^+\rightarrow J/\psi\eta K^+$ decay}
\begin{document}
\maketitle
\begin{abstract}

Recently, a study on the $J/\psi\eta$ mass spectrum from
$B^+\rightarrow J/\psi\eta K^+$ decays was reported by the LHCb
detector. The results of this study are reported as a ratio of
branching fractions as
$F_{X}\equiv\frac{\mathcal{B}r(B^+\rightarrow
XK^+)\times\mathcal{B}r(X\rightarrow
J/\psi\eta)}{\mathcal{B}r(B^+\rightarrow \psi(2S)
K^+)\times\mathcal{B}r(\psi(2S)\rightarrow J/\psi\eta)}$ for
$X=\psi_2(3823),\psi(4040)$, which are
$(5.95^{+3.38}_{-2.55})\times10^{-2}$ and
$(40.60\pm11.20)\times10^{-2}$, respectively. Also, the products
related to $B_{X}\equiv\mathcal{B}r(B^+\rightarrow
XK^+)\times\mathcal{B}r(X\rightarrow J/\psi\eta)$ branching
fractions are
$B_{\psi_2(3823)}=(1.25^{+0.71}_{-0.53}\pm0.04)\times10^{-6}$ and
$B_{\psi(4040)}=(8.53\pm2.35\pm0.30)\times10^{-6}$. For the first
time, we calculated this branching fraction using factorization.
According to our calculations, $F_X$ to be
$F_{\psi_{2}(3823)}=(6.55\pm1.88)\times10^{-2}$ and
$F_{\psi(4040)}=(14.33\pm4.15)\times10^{-2}$ at $\mu=m_b/2$. We
have estimated $B_{\psi_{2}(3823)}=(0.26\pm0.05)\times10^{-6}$ at
$\mu=m_b/2$ and $B_{\psi(4040)}=(2.88\pm0.64)\times10^{-6}$ at
$\mu=2m_b$.
\end{abstract}

\section{Introduction}
The decay modes of $B$ mesons has provided a good opportunity to
probe the physics beyond the standard model (SM), studies of
charmonium and to search over the properties of new
charmonium-like exotic particles beyond this model.\\
In general, studies of various hadronic transitions in charmonia
parts can clarify the internal structure of particles where the
transition is accompanied by the emission of an $\eta$ meson (in
$e^+e^-\rightarrow J/\psi\eta$ processes) \cite{BESIII1}.\\
This structure is largely unknown for newly discovered hadronic
states. In the last two decades, a large number of new hadron states have been discovered, including $\chi_{c1}(3872)$ in the decay of the beauty hadrons into charmonia \cite{Belle1}, \cite{N.B1}.\\
Over the years, various charmonium states have been reported
experimentally, including the hidden charm structures
$\psi_2(3823)$ \cite{BESIII2} and $X(3960)$. The discovery of this
meson aroused the interest of elementary particle physics
researchers. More information about the internal structure and the
dynamics of its creation can be extracted by interpreting the
types of weak decay and mesons. Theoretically, an enormous effort
has been made to reveal the nature of unexpected exotic hadrons
using different approaches \cite{XKD1}.\\
The description of the basic dynamics for non-leptonic decays of
the $B$ meson is very complicated, but for systems with heavy $b$-
and $c$-quarks, it is greatly simplified due to the factorization
of the hadronic matrix elements in terms of decay constants and
form
factors \cite{C.A1}.\\
In \cite{CH.W} and \cite{MB}, they have obtained a factorization
formula for heavy final states, which includes elements of the
naive factorization approach and the hard scattering approach.
This is an accurate basis for factorizing the non-leptonic
two-body $B$ meson decay in the heavy quark limit. Due to the
intrinsic scale of strong interactions, non-leptonic decay
amplitudes will have a simple structure in the heavy-quark limit
\cite{CH.W}. If the mass of the decaying weak quark is large
enough, there will be many predictions for $B$ decay with
$CP$-violation in the heavy-quark limit. This makes available the
amplitudes of these decays in terms of experimentally measurable
semi-leptonic form factors, hadronic light-cone distribution
amplitudes, and hard-scattering functions calculable in
perturbative QCD \cite{MB}. Naive factorization of four fermion
operators for many (but not all) non-leptonic decays means that
so-called "non-factorizable" corrections, hitherto thought to be
intractable, can be accurately calculated.\\
Therefore, with the naive factorization approach, which is a
common method for solving the non-leptonic hadronic decays of $B$
into charmed modes, these decays are calculated \cite{M.W1}. Also,
for a system with heavy $c$-quark, the non-relativistic method can be used in calculations.\\
In charmed decay processes, the contribution of penguin operators
is negligible due to the existence of $c$-quark in calculations.
This leads to negligible theoretical uncertainties in the relevant
quantum chromodynamics (QCD) dynamics.\\
The S-wave Low-lying modes with full charm are systematically
investigated considering diquark-diquark-antiquark configurations.
In addition, all possible color structures in each configuration
are considered, along with their couplings. The total angular
momentum $J$, which corresponds to the total spin $S$, ranges from
1/2 to 2/5 with negative parity $P=-1$ \cite{GY}. The
$B^+\rightarrow XK^+$ decay is achieved through the weak
$\bar{b}\rightarrow\bar{c}cs$ decay over very short distances \cite{E.B2}.\\
Low energy QCD remains a field of high interest both
experimentally and theoretically. Recent studies have considered
$\psi_2$(3823) as a good candidate for spin-triplet members.
Experimental information on the $\psi_2$(3823) is still sparse. In
this paper, it is to provide additional theoretical predictions
for the correct assignment of the $\psi_2$(3823) to be the $J=2$
spin-triplet partner, by comparing decay channels to the
experimental measurements \cite{MA}. Recently, LHCb collaboration
have obtained results for masses of the $X$ state
($X=\psi_2(3823),\psi(4040)$) and charmonium states and measured
the $F_X$ ratio and the product of the branching fractions $B_X$.
They measured an upper limit at 90\% CL on the ratio of branching
fractions $F_X(m_X)$ (refer to Eq. \ref{eq9}) for
$B^+\rightarrow(X\rightarrow J/\psi\eta)K^+$ through a narrow
intermediate $X$ state for particle mass $X$ \cite{LHcb1}.\\
The ratios of branching fractions are reported to be
$F_{\psi_2(3823)}=(5.95^{+3.38}_{-2.55})\times10^{-2}$ and
$F_{\psi(4040)}=(40.60\pm11.20)\times10^{-2}$. Also, the
corresponding products of branching fractions are
$B_{\psi_2(3823)}=(1.25^{+0.71}_{-0.53}\pm0.04)\times10^{-6}$ and
$B_{\psi(4040)}=(8.53\pm2.35\pm0.30)\times10^{-6}$
\cite{LHcb1}.\\
In this study, we have calculated the branching fractions for the
$B^+\rightarrow\psi(2S)K^+$, $\psi(2S)\rightarrow J/\psi\eta$,
$B^+\rightarrow XK^+$ and $X\rightarrow J/\psi K^+$ decays under
the factorization approach and obtained
$F_{\psi_2(3823)}=(6.55\pm1.88)\times10^{-2}$ and
$F_{\psi(4040)}=(14.33\pm4.15)\times10^{-2}$ at $\mu=m_b/2$. We
have estimated $B_{\psi_{2}(3823)}=(0.26\pm0.05)\times10^{-6}$ at
$\mu=m_b/2$ and $B_{\psi(4040)}=(2.88\pm0.64)\times10^{-6}$ at
$\mu=2m_b$.

\section{Decay amplitudes, decay rates and branching fractions}
\subsection{The $B^+\rightarrow \psi_2(3823)K^+$, $B^+\rightarrow\psi(4040)K^+$ and $B^+\rightarrow\psi(2S)K^+$ decays}

\begin{figure}[t]
\begin{center} \includegraphics[scale=0.9]{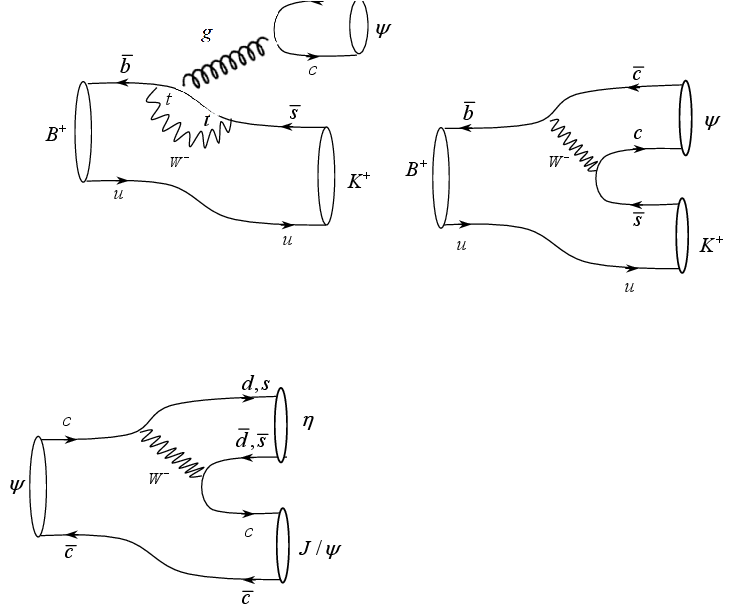}
\caption{\label{fig1}Feynman's diagrams contributing to
$B^+\rightarrow\psi,\psi_2 K^+$ and $\psi,\psi_2\rightarrow
J/\psi\eta$ decays.}
\end{center}
\end{figure}

There are various phenomenological methods to investigate the
properties of different decays of the $B$ meson. Factorization has
long been a noteworthy idea in hadronic decays of heavy mesons.
The separation of scales associated with the tiny binding energy
of the $X$ makes this an ideal system for applying effective field
theory. A field theory with contact interactions between the charm
mesons is a good first approximation provided one takes into
account the existence of inelastic scattering channels for the
charm mesons. The effective field theory can also be of use by
removing scales of negligible effect from calculations and
simplifying the remaining calculations, by expanding in the ratios
of the scales present in the problem. Here we are going to discuss
one application of effective field theory in the study of meson
decay. In the phenomenological behavior of weak hadrons decays,
the starting point is the effective weak Hamiltonian at low
energy. For the $B^+\rightarrow\psi K^+$ ($\psi: \psi(2S),
\psi(4040)$) and $B^+\rightarrow\psi_2(3823)K^+$ decays, according
to Fig.\ref{fig1}, the decay amplitude can be written as follows
\cite{B.M33}
\begin{eqnarray}\label{eq1}
\mathcal{M}(B^+\rightarrow\psi
K^+)=i\sqrt{2}G_{F}m_{\psi}f_{\psi}(\epsilon\cdot p_B)
f_+^{B^+\rightarrow
K^+}(m_{\psi}^2)\Big(a_2V_{cb}V_{cs}^*-V_{tb}V_{ts}^*(a_3+a_9+
r_\chi^{\psi}(a_5+a_7))\Big),
\end{eqnarray}
and \cite{HAI}
\begin{eqnarray}\label{eq2}
\mathcal{M}(B^+\rightarrow\psi_2K^+)=i\sqrt{2}G_{F}m_Bf_{\psi_2}p_c
f_+^{B^+\rightarrow
K^+}(m_{\psi_2}^2)\Big(a_2V_{cb}V_{cs}^*-V_{tb}V_{ts}^*(a_3+a_9+
r_\chi^{\psi_2}(a_5+a_7))\Big),
\end{eqnarray}
where the $f_{\psi_2}=1/m_{\psi_2}g_{\psi_2J/\psi\eta}$ \cite{EK},
$f_{\psi}$ and $f_{\eta}$ are decay constant and
$r_\chi^{\psi,\psi_2}=(2m_{\psi,\psi_2}/m_b)(f_{\psi,\psi_2}^{\bot}/f_{\psi,\psi_2})$
\cite{D.Me}. To achieve the form factor $f_+^{B^+\rightarrow
K^+}(m_{\psi,\psi_2}^2)$ we take the form \cite{WGP}
\begin{eqnarray}\label{eq3}
f_+(q^2)&=&\frac{\mathcal{L}}{(1-q^2/m_{B^*_s}^2)}\sum\limits_{n=0}^{N-1}a^+_n\Big(z^n-\frac{n}{N}(-1)^{n-N}z^N\Big),\\
z(q^2,t_0)&=&\frac{\sqrt{t_+-q^2}-\sqrt{t_+-t_0}}{\sqrt{t_+-q^2}+\sqrt{t_+-t_0}},
\end{eqnarray}
here $N=3$, $t_+=(m_{B^+}+m_{K^+})^2$ and $t_0=0$ (it means
$z(q^2=0)=0$). $m_{B^*_s}$ is the pole mass. The values of the fit
coefficients $a^+_n$ and the chiral logarithm factor $\mathcal{L}$
are \cite{WGP}
\begin{eqnarray}\label{eq4}
a^+_0=0.2545(90),\quad a^+_1=-0.71(14),\quad
a^+_2=0.32(59),\quad\mathcal{L}=1.304(10).
\end{eqnarray}
Also, $G_F=(1.16639\pm0.00001)\times 10^{-5} GeV^{-2}$,
$V_{pb}V_{ps}^* (p= c,t)$ are the Cabibbo-Kobayashi-Maskawa (CKM)
matrix elements \cite{PDG1} and the quantities $a_i$ (i = 1,
\ldots, 10) are the following combinations of the effective Wilson
coefficients
\begin{eqnarray}\label{eq4}
a_{2j-1}=c_{2j-1}+\frac{1}{3}c_{2j},\quad
a_{2j}=c_{2j}+\frac{1}{3}c_{2j-1},\quad j=1,2,3,4,5.
\end{eqnarray}
The Wilson coefficients, $c_j$, in the effective weak Hamiltonian
have been reliably evaluated by the next-to-leading logarithmic
order. To proceed, we use the following numerical values at three
different choices of $\mu= 2m_b, m_b, m_b/2$ scales \cite{M.Be1},
which have been obtained in the next-to-leading logarithm in the
naive dimensional regularization (NDR) scheme that are shown in
Tab. \ref{tab1}.\\
\begin{table}[t]
\centering\caption{\label{tab1}  Wilson coefficients $c_j$ in the
NDR scheme ($\alpha=1/129$).}
\begin{tabular}{|c||c|c|c|}
  \hline
   NLO    & $\mu=m_b/2$ & $\mu=m_b$ & $\mu=2m_b$\\ \hline\hline
 $c_1$ & 1.137 & 1.081 & 1.045 \\ \hline
 $c_2$ & -0.295 & -0.190 & -0.113 \\ \hline
 $c_3$ & 0.021 & 0.014 & 0.009 \\ \hline
 $c_4$ & -0.051 & -0.036 & -0.025 \\ \hline
 $c_5$ & 0.010 & 0.009 & 0.007 \\ \hline
 $c_6$ & -0.065 & -0.042 & -0.027 \\ \hline
 $c_7/\alpha$ & -0.024 & -0.011 & 0.011 \\ \hline
 $c_8/\alpha$ & 0.096 & 0.060 & 0.039 \\ \hline
 $c_9/\alpha$ & -1.325 & -1.254 & -1.195 \\ \hline
 $c_{10}/\alpha$ & 0.331 & 0.223 & 0.144 \\ \hline
\end{tabular}
\end{table}
The branching fraction the $B^+\rightarrow \psi_2(3823)K^+$ and
$B^+\rightarrow\psi K^+$ decays under the factorization approach
is given by \cite{HAI}, \cite{M.E1}
\begin{eqnarray}\label{eq5}
\mathcal{B}r(B^+\rightarrow \psi_2,\psi
K^+)&=&\frac{|\vec{p_c}|}{\Gamma_{tot}}\frac{|\mathcal{M}
(B^+\rightarrow \psi_2,\psi K^+)|^2}{8\pi m_B^2},
\end{eqnarray}
where the $\Gamma_{tot}$ for charged $B$ meson is
$(4.02\pm0.01)\times10^{-13}$ GeV \cite{PDG1} and and
$|\vec{p}_c|$ is the absolute value of the 3-momentum of the
$\psi, \psi_2$ (or the $K^+$) in the rest frame of the $B^+$
meson.

\subsection{The $\psi(2S)\rightarrow J/\psi\eta$, $\psi(4040)\rightarrow J/\psi\eta$ and $\psi_2(3823)\rightarrow J/\psi\eta$ decays}

It is clear that the physics of charmonium is not yet fully
resolved and the field is full of challenges and opportunities.
Due to the advancement of techniques and devices, detection
accuracy has increased greatly in the last decade. As a result,
new puzzles are continuously created.\\
In general, we can divide the strong $\psi(3823)$ and $\psi(4040)$
decays modes into three types: open charm decay, hidden charm
decay ($J/\psi f$ where $f$ are light mesons) \cite{GX} and decay
into light-hadrons (L-H decay) \cite{XLZ1}. The $J/\psi P$ and L-H
decay processes are not suppressed by Okubo-Zweig-Izuka(OZI) \cite{GLQ}.\\
The $\psi\rightarrow J/\psi\eta$ and $\psi_2\rightarrow
J/\psi\eta$ decays are $V\rightarrow VP$ and $T\rightarrow VP$
type decays where $V, P, T$ are vector, pseudoscalar and tensor
mesons respectively. To calculate the $\psi(4040)\rightarrow
J/\psi\eta$ and $\psi(2S)\rightarrow J/\psi\eta$ decays we have
\cite{M.A.1}
\begin{eqnarray}\label{eq5}
\Gamma(\psi\rightarrow
J/\psi\eta)=\frac{g^2}{96\pi}\frac{\lambda^{3/2}(m^2_{\psi},m^2_{\eta},m^2_{J/\psi})}{m^3_{\psi}},
\end{eqnarray}
where $\lambda(x, y, z)=x^2+y^2+z^2-2(xy+yz+zx)$ and the coupling
$g=(0.218\pm0.003) GeV^{-1}$ \cite{Qi.H}, \cite{N.N.}.\\
For the $\psi_2(3823)\rightarrow J/\psi\eta$ we have \cite{FG}
\begin{eqnarray}\label{eq9}
\Gamma(\psi_2(3823)\rightarrow
J/\psi\eta)=\alpha_{\psi_2J/\psi\eta}\frac{P^5_{\psi_2J/\psi\eta}}{10\pi}g^2_{\psi_2J/\psi\eta},
\end{eqnarray}
where $P_{\psi_2J/\psi\eta}= \lambda^{1/2}(m^2_{\psi_2},
m^2_{J/\psi}, m^2_{\eta})/2m_{\psi_2}$ is the three-momentum of
the final states (pseudoscalar mesons) in the rest frame of the
decaying initial state (tensor meson) and
$\lambda(x,y,z)=x^2+y^2+z^2-2xy-2xz-2yz$ is the K\"{a}llen
triangle function. The $\alpha_{\psi_2J/\psi\eta}$ takes into
account the average over spin of the initial state and the sum
over final isospin states with averaging over initial isospin
states \cite{FG}, $g_{\psi_2J/\psi\eta}$ is the
effective $\psi_2(3823)\rightarrow J/\psi\eta$ coupling constant which we get 22.22.\\
We calculated the branching fraction $\psi_2(3823)\rightarrow
J/\psi\eta$, $\psi(4040)\rightarrow J/\psi\eta$ and
$\psi(2S)\rightarrow J/\psi\eta$ decays. The $\Gamma_{tot}$ for
$\psi(3823)$, $\psi(4040)$ and $\psi(2S)$ meson are $\Gamma<2.9$,
$84\pm12$ and $(293\pm9)\times10^{-3}$ MeV respectively
\cite{PDG1}.

\section{The ratio and product of branching fractions:\\ $F_X$ and $B_X$}

The LHCb collaboration, a study of the $J/\psi\eta$ mass spectrum
from $B^+\rightarrow J/\psi K^+$ decays, have reported. They
discussed an exceptional feature of the $X$ extensively. Their
results have been obtained in the form of a ratio of branching
fractions using the $B^+\rightarrow(\psi(2S)\rightarrow
J/\psi\eta) K^+$ decay. The $F_X$
ratio and the product of branching fractions $B_X$ have been obtained with the following formulas \cite{LHcb1}: \\
\begin{eqnarray}\label{eq10}
&&F_{X}\equiv\frac{\mathcal{B}r(B^+\rightarrow
XK^+)\times\mathcal{B}r(X\rightarrow
J/\psi\eta)}{\mathcal{B}r(B^+\rightarrow\psi(2S)
K^+)\times\mathcal{B}r(\psi(2S)\rightarrow J/\psi\eta)}, \nonumber\\
&&B_{X}\equiv\mathcal{B}r(B^+\rightarrow
XK^+)\times\mathcal{B}r(X\rightarrow J/\psi\eta)
\end{eqnarray}
We calculated the value of $F_X$ and $B_X$ using the branching
ratio that we estimated. We obtained the value of $F_X$ in three
different scales and calculated the value of $B_X$ using it. In
process related to $B^+\rightarrow XK^+$, $X$ can be formed as an
intermediate resonance particle and subsequently it can decay into
$J/\psi\eta$. Therefore, $B_X$, is related to meson $X$ and its
two-body decay ($X\rightarrow J/\psi\eta$).

\section{Numerical results and conclusion}
The meson masses and decay constants are tabulated in Tab.
\ref{tab2}.
\begin{table}[t]
\centering\caption{\label{tab2} The meson masses and decay
constants (in MeV) \cite{PDG1}}
\begin{tabular}{|c c c c c|}
  \hline
   $m_{\psi_2(3823)}$ & $m_{\psi(4040)}$ & $m_{\psi(2S)}$ & $m_{K^+}$ & $m_{\eta}$
\\\hline
$3823.51\pm0.34$ & $ 4040\pm4$ & $3686.097\pm0.011$ &
$493.677\pm0.015$ & $547.862\pm0.017$
\\\hline
 $m_{B^+}$ & $m_{B_s^*}$ & $m_{J/\psi}$ & $m_{b}$ & \\\hline
 $5279.41\pm0.07$ & $5415.4\pm1.4$ & $3096.900\pm0.006$ &
$4183\pm7$ &
\\\hline\hline
 $f_{\psi(2S)}$ & $f_{\psi(2S)}^{\bot}$ \cite{F.M},\cite{P.B} & $f_{\eta}$ \cite{B.M3} & $f_{\psi(4040)}$ & $f_{J/\psi}^{\bot}$\\\hline
 $282\pm14$ & $255\pm33$ & $131\pm7$ & $319\pm22$ \cite{Guo} & $405\pm0.014$\\\hline
\end{tabular}
\end{table}
The elements of the CKM matrix used in the calculations have the
following values:\\
\begin{tabular}{ccccc}
$|V_{cs}|$ & $|V_{cb}|$ & & $|V_{tb}|$ & $|V_{ts}|$\\
$0.975\pm0.006$ & $(41.1\pm1.2)\times10^{-3}$ & & $1.010\pm0.027$ & $(41.5\pm0.9)\times10^{-3}$ \\
\end{tabular}\\
Charmonium decays are ideal for studying as an investigation. We
calculated the $\psi\rightarrow J/\psi\eta$,
$\psi_2(3823)\rightarrow J/\psi\eta$,$B^+\rightarrow \psi K^+$ and
$B^+\rightarrow \psi_2(3823)K^+$ decays by drawing the Feynman
diagrams related to these decays and determining their
contributions using the naive factorization approach, decay
amplitude and the rate of decays and, finally, the branching
ratios. The numerical result of branching ratio for the
$\psi(4040)\rightarrow J/\psi\eta$, $\psi_2(3823)\rightarrow
J/\psi\eta$ and $\psi(2S)\rightarrow J/\psi\eta$ decays is listed
in Tab. \ref{tab3}.
\begin{table}[t]
\centering\caption{\label{tab3} The numerical result of branching
ratio for the charmonium decays.}
\begin{tabular}{|c c c c c|}
  \hline
  Decay mode &  & This work &  & Exp. $(\times10^{-2})$ \cite{PDG1} \\
\hline $\mathcal{B}r(\psi_2(3823)\rightarrow J/\psi\eta)$ &
 & $10.98\pm0.76$ &  &
$<14$ \\\hline $\mathcal{B}r(\psi(4040)\rightarrow J/\psi\eta)$ &
& $0.48\pm0.06$ &  & $0.52\pm0.07$
\\\hline $\mathcal{B}r(\psi(2S)\rightarrow J/\psi\eta)$ &
 & $3.35\pm0.08$ &  & $3.37\pm0.05$
\\\hline
\end{tabular}
\end{table}
The results of branching ratio for the
$B^+\rightarrow\psi(4040)K^+$,
$B^+\rightarrow\psi(2S)K^+$ and $B^+\rightarrow\psi_2(3823)K^+$ by comparing the experimental value are presented in Tab. \ref{tab4}.\\
\begin{table}[t]
\centering\caption{\label{tab4} The numerical result of branching
ratio for the $B^+\rightarrow\psi(2S)K^+$ (in units of $10^{-4}$),
$B^+\rightarrow\psi(4040)K^+$ (in units of $10^{-3}$) and
$B^+\rightarrow\psi_2(3823)K^+$ (in units of $10^{-6}$) decays,
the ratio of branching fractions $F_X$ (in units of $10^{-2}$) and
product of branching fractions $B_X$ (in units of $10^{-6}$).}
\begin{tabular}{|ccccc|}
  \hline
 Decay mode & $\mu=m_b/2$ & $\mu=m_b$ & $\mu=2m_b$ & Exp.  \\
\hline $\mathcal{B}r(B^+\rightarrow\psi(2S)K^+)$ & $1.20\pm0.24$ &
$3.50\pm0.66$ & $6.17\pm1.11$ & $6.24\pm0.20$ \cite{PDG1}\\ \hline
 $\mathcal{B}r(B^+\rightarrow \psi(4040)K^+)$ &
$0.12\pm0.02$ & $0.34\pm0.06$ & $0.60\pm0.11$ & $1.60\pm0.50$
\cite{PDG1}
\\\hline
$\mathcal{B}r(B^+\rightarrow \psi_2(3823)K^+)$ & $2.40\pm0.46$ &
$1.08\pm0.24$ & $0.75\pm0.16$ & $1.20\pm0.60$ \cite{PDG1}
\\\hline\hline
$F_{\psi_2(3823)}$ & $6.55\pm1.88$ & $1.01\pm0.30$ & $0.40\pm0.11$
& $5.95^{+3.38}_{-2.55}$ \cite{LHcb1}
\\\hline $F_{\psi(4040)}$ & $14.33\pm4.15$ & $13.92\pm4.00$ &
$13.93\pm3.99$ & $40.60\pm11.20$ \cite{LHcb1} \\\hline\hline
$B_{\psi_2(3823)}$ & $0.26\pm0.05$ & $0.12\pm0.03$ & $0.08\pm0.02$
& $1.25^{+0.71}_{-0.51}\pm0.04$ \cite{LHcb1}
\\\hline $B_{\psi(4040)}$ & $0.58\pm0.12$ & $1.63\pm0.35$ &
$2.88\pm0.64$ & $8.53\pm2.35\pm0.30$ \cite{LHcb1} \\\hline
\end{tabular}
\end{table}
We calculated the $B^+\rightarrow \psi_2(3823)K^+$,
$B^+\rightarrow\psi(4040)K^+$ and $B^+\rightarrow \psi(2S)K^+$
decay in three scales of $\mu=m_b$ and in this way we obtained the
values of $F_X$ and $B_X$ in these scales. The
$\mathcal{B}r(B^+\rightarrow\psi(4040)K^+)$ and
$\mathcal{B}r(B^+\rightarrow \psi(2S)K^+)$ at $\mu=2m_b$, and
$\mathcal{B}r(B^+\rightarrow \psi_2(3823)K^+)$ at $\mu=m_b$ are in good agreement with experimental results.\\
Also, this decay is one of the strong decays and is of the type of
decay of charm hadron meson into two light heavy mesons. The
amazing $\psi$ meson has a heavy mass but the properties of light
mesons. This method has been used for the
$\mathcal{B}r(\psi\rightarrow J/\psi\eta)$, and the
results consistent with the experimental data were obtained.\\
With recent investigations of $B$ factories and the planned
emphasis on heavy flavor physics in future experiments, the role
of $B$ decay in providing fundamental tests of the standard model
and potential effects of new physics will continue to grow.\\
In this study, we examined the decay properties of the tensor
meson $\psi_2(3823)$. The experimental value of the decay
branching ratio $B^+\rightarrow \psi_2(3823)K^+$ has not yet been
measured, but based on our recent composite data analysis
\cite{LHcb1}, we calculated it. The exotic states such as
$\psi(4040)$ are identified as the low-lying di-mesonic molecular
states in the charm sector and largely qualify as $c\bar{c}$
state. The branching fraction of the $\psi_2(3823)\rightarrow
J/\psi\eta$ decay is larger than that of the $\psi(2S)\rightarrow
J/\psi\eta$ decay. Because of higher charmonium excitations, the
charmonium-to-charmonium transitions are not suppressed by $\eta$
meson emission. This also applies to the $B^+\rightarrow
\psi_2(3823)K^+$ decay and, according to Tab. \ref{tab4}, the
increasing trend of the value of the branching ratio is the
opposite of the two $B^+\rightarrow\psi(4040)K^+$ and
$B^+\rightarrow \psi(2S)K^+$ decays.\\ Our work provides a precise
framework for evaluating strong interactions for a large
classification of two-body non-leptonic $B$ decays. In the case of
branching ratios, theoretical predictions have large uncertainties
due to the hadronic distributions, the hard scattering, and the
renormalization scales in factorizable amplitudes.\\
Uncertainties are caused by our inability to perform infinitely
precise calculations. These uncertainties are not determined by
stochastic processes, since they lead to the same results if
repeated with the same assumptions. Choosing between two temporal
models, such a distribution has a limited ability to
probabilistically enclose the space of possibilities. In other
cases, such as uncertainties arising from scaling calculations,
the interpretation of such distributions is uncertain at best. For
example, by changing the scale from $m_b$ to $m_b/2$, the results
decrease and this uncertainty is also present in our
calculations.\\
In the heavy quark limit, matrix elements can be
expressed in terms of certain non-perturbative input values such
as transition form factors. The errors appearing in the results
belong to the uncertainties in the input parameters and the
systematic uncertainties.\\ The power corrections beyond the heavy
quark limit generally introduce large theoretical uncertainties.
Also, the CKM factors mainly provide a general factor for the
branching ratios and do not introduce many uncertainties to the
numerical results.


\begin{thebibliography}{20}
\bibitem{BESIII1}
M. Ablikim et al., BESIII collaboration, \textit{Observation of
the Y(4220) and Y(4360) in the process $e^+e^-\rightarrow\eta
J/\psi$}, Phys. Rev. D \textbf{102} (2020) 031101.
\bibitem{Belle1}
S.K. Choi et al., Belle collaboration, \textit{Observation of a
narrow charmoniumlike state in exclusive $B^{\pm}\rightarrow
K^{\pm}\pi^+p^-J/\psi$ decays}, Phys. Rev. Lett \textbf{91} (2003)
262001.
\bibitem{N.B1}
N. Brambilla et al., \textit{The XYZ states: experimental and
theoretical status and perspectives}, Phys. Rept \textbf{873}
(2020) 1.
\bibitem{BESIII2}
M. Ablikim et al., BESIII collaboration, \textit{Observation of
resonance structures in $e^+e^-\rightarrow\pi^+\pi^-\psi_2(3823)$
and mass measurement of $\psi_2(3823)$}, Phys. Rev. Lett
\textbf{129} (2022) 102003.
\bibitem{XKD1}
X.K. Dong, F.K. Guo and B.S. Zou, \textit{Explaining the many
threshold structures in the heavy-quark hadron spectrum}, Phys.
Rev. Lett \textbf{126} (2021) 152001.
\bibitem{C.A1}
C.A. Morales, N. Quintero, C.E. Vera and A. Villalba,
\textit{Analysis of the nonleptonic charmonium modes
$B^0_s\rightarrow j/\psi f^{'}_2(1525)$ and $B^0_s\rightarrow
j/\psi K^+K^-$}, Phys. Rev. D \textbf{95} (2017) 036013.
\bibitem{CH.W}
CH.W. Bauer, D. Pirjol, and I.W.
Stewart, \textit{A proof of Factorization for $B\rightarrow
D\pi$}, Phys. Rev. Lett. \textbf{87} (2001) 201806.
\bibitem{MB}
M. Beneke, G. Buchalla, M. Neubert and C.T. Sachrajda, \textit{QCD
factorization for exclusive non-leptonic B-meson decays: General
arguments and the case of heavy-light final states}, Nucl. Phys. B
\textbf{591} (2000) 313.
\bibitem{M.W1}
M. Wirbel, B. Stech and M. Bauer, \textit{Exclusive semileptonic
decays of heavy mesons}, Z. Phys. C \textbf{29} (1985) 637.
\bibitem{GY} G. Yang, J. Ping and J. Segovia,
\textit{Fully-charm and -bottom pentaquarks in a Lattice-QCD
inspired quark model}, Phys. Rev. D \textbf{106} (2022) 014005.
\bibitem{E.B2}
E. Braaten and M. Kusunoki, \textit{Exclusive production of the
$X(3872)$ in $B$ meson decay}, Phys. Rev. Lett \textbf{71} (2005)
074005.
\bibitem{MA}
M. Ablikim et al., BESIII collaboration, \textit{Search for new
decay modes of the $\psi_2$(3823) and the process
$e^+e^-\rightarrow \pi^0\pi^0\psi_2(3823)$}, Phys. Rev. D
\textbf{103} (2021) L091102.
\bibitem{LHcb1}
R. Aaij et al., LHCb collaboration, \textit{Study of charmonium
and charmonium-like contributions in $B^+\rightarrow J/\psi\eta
K^+$ decays}, J. High. Energ. Phys \textbf{46} (2022).
\bibitem{B.M33} B. Mohammadi, \textit{Estimating
of CP-violation in decay}, Phys. Part. Nucl. Lettr \textbf{14}
(2017) 886.
\bibitem{HAI}
H.Y. Cheng, CH.W. Chiang and ZH.Q.
Zhang, \textit{Two- and three-body hadronic decays of charmed
mesons involving a tensor meson}, Phys. Rev. D \textbf{105} (2022)
093006.
\bibitem{EK} E. Katz, A. Lewandowski and M.D. Schwartz,
\textit{Tensor mesons in AdS/QCD}, Phys. Rev. D \textbf{74} (2006)
086004.
\bibitem{D.Me}
D. Melikhov and B. Stech, \textit{Weak form factors for heavy
meson decays: An update}, Phys. Rev. D \textbf{62} (2000) 014006.
\bibitem{WGP} W.G. Parrott, C. Bouchard and
C.T.H. Davies (HPQCD collaboration), \textit{$B\rightarrow K$ and
$D\rightarrow K$ form factors from fully relativistic lattice
QCD}, Phys. Rev. D \textbf{107} (2023) 014510.
\bibitem{PDG1}
S. Navas et al., Particle Data Group, \textit{Review of particle
physics}, Phys. Rev. D \textbf{110} (2024) 030001.
\bibitem{M.Be1}
M. Beneke, G. Buchalla, M. Neubert and C.T. Sachrajda, \textit{QCD
factorization in $B\rightarrow \pi K, \pi\pi$ decays and
extraction of wolfenstein parameters}, Nucl. Phys. B \textbf{606}
(2001) 245.
\bibitem{M.E1} B. Mohammadi and E. Amirkhanlou,
\textit{Charge parity violation parameters in neutral beauty meson
decays into charmonium and strange kaon mesons}, Int. J. Mod.
Phys. A \textbf{37} (2022) 2250082.
\bibitem{GX}
G. Li and X.H. Liu, \textit{Investigating possible decay modes of
$Y(4260)$ under the $D_1(2420)\bar{D}+c.c$ molecular state
ansatz}, Phys. Rve. D \textbf{88} (2013) 094008.
\bibitem{XLZ1}
X. Liu, B. Zhang and X.Q. Li, \textit{The puzzle of excessive
non-$D\bar{D}$ component of the inclusive $\psi(3770)$ decay and
the long-distant contribution}, Phys. Lettr. B \textbf{675} (2009)
441.
\bibitem{GLQ}
G Li, Qiang Zhao and Chao-Hsi Chang, \textit{Decays of $J/\psi$
and $\psi'$ into vector and pseudoscalar meson and the
pseudoscalar glueball-$q\bar{q}$ mixing}, J. Phys. G \textbf{35}
(2008) 055002.
\bibitem{M.A.1}
M. Albaladejo, J.T. Daub, C. Hanhart, B. Kubis and B. Moussallam,
\textit{How to employ $\bar{B}^0_d\rightarrow J/\psi(\pi\eta,
K\bar{K}$) decays to extract information on $\pi\eta$ scattering},
JHEP \textbf{04} (2017) 010.
\bibitem{Qi.H}
Q. Huang and J.J. Wu, \textit{Proposal to detect a moving triangle
singularity in $\psi(2S)\rightarrow\pi^+\pi^-K^+K^-$ process},
Phys. Rev. D \textbf{104} (2021) 116003.
\bibitem{N.N.} N.N. Achasov and G.N. Shestakov, \textit{Description of the
$\psi$(3770) resonance interfering with the background}, Phys.
Rev. D \textbf{87} (2017) 057502.
\bibitem{FG} F. Giacosa, Th. Gutsche, V.E.
Lyubovitskij and Amand Faessler, \textit{Decays of tensor mesons
and the tensor glueball in an effective field approach}, Phys.
Rev. D \textbf{72} (2005) 114021.
\bibitem{F.M}
F.M. Al-Shamali and A.N. Kamal, \textit{Nonfactorization and final
state interactions in $(B,B_s)\rightarrow\psi P$ and $\psi V$
decays}, Eur. Phys. J. C \textbf{4} (1998) 669.
\bibitem{P.B}
P. Ball and V.M. Braun, \textit{The $\rho$ meson light-cone
distribution amplitudes of leading twist revisited}, Phys. Rev. D
\textbf{54} (1996) 2182.
\bibitem{B.M3}
B. Mohammadi and H. Mehraban, \textit{Studies of three-body decay
of $B$ to $J/\psi\eta K$ and $B(B_S)$ to $\eta_c\pi K^*$}, Adv.
High. Ener. Phys \textbf{2014} (2014) 451613.
\bibitem{Guo} G.L. Wang, \textit{Decay constants
of heavy vector mesons in relativistic bethe-salpeter method},
Phys. Lett. B \textbf{633} (2006) 492.
\end{thebibliography}
\end{document}